\documentstyle[epsfig,aps,pra]{revtex}
\begin{document}

\newcommand{\ket}[1]{| #1 \rangle}
\newcommand{\bra}[1]{\langle #1 |}
\newcommand{\braket}[2]{\langle #1 | #2 \rangle}
\newcommand{\proj}[1]{| #1\rangle\!\langle #1 |}

\twocolumn[\hsize\textwidth\columnwidth\hsize\csname
@twocolumnfalse\endcsname

\title{Comment on: ``A quantum approach to static games of complete information''} 

\author{S. C. Benjamin}

\address{
Centre for Quantum Computation, Clarendon Laboratory, University of Oxford, OX1 3PU, UK \\
Email: {\tt s.benjamin@qubit.org}
}  

\maketitle
\bigskip

]

In a recent article \cite{origArt}, an interesting new scheme for quantizing games was introduced, and
the scheme was applied to the famous game {\em Battle of the Sexes}. In its traditional form, this game
has two equivalent stable solutions (Nash equilibria), and the players face a dilemma in choosing
between them. One conclusion of the article was that the dilemma is removed in the quantum game, i.e. it
has a unique solution. In this Comment we make two observations. Firstly, the overall
quantization scheme is fundamentally very similar to a previous scheme proposed by Eisert {\em et al}
\cite{eisertPRL,EisertJMO} -- the similarity is non-obvious because of the very different use of the
word `strategy' in the two approaches. Secondly, we argue that the quantum {\em Battle
of the Sexes} game does not in fact have a unique solution, hence the players {\em are}
still subject to a dilemma, although it may be easier to resolve than in the traditional game.

The proposed scheme for quantizing two-player static games is shown in Fig 1a(i). Marinatto and
Weber use the word `strategy' to refer to a state $S$ in a $2\otimes 2$ dimensional Hilbert space.
This `strategy' is actually supplied to the players initially, they then manipulate it
during a `tactics' phase, and finally it is measured. The result of the measurement directly determines
the payoff to the players
\cite{theSame}, as shown in Fig. 1a(ii). In Ref.
\cite{origArt} the labels $\ket{O}$ and $\ket{T}$ are used for the measurement basis, whilst in Ref.
\cite{eisertPRL} it is $\ket{C}$ and $\ket{D}$; here we will use the generic labels $\ket{0}$ and
$\ket{1}$. During the `tactics' phase of Marinatto and
Weber's scheme, each player may only manipulate $S$ within her own $2$
dimensional subspace, and the measurement basis respects this division. Hence it is useful to
think of
$S$ as being the global state of a pair of two-state quantum systems or `qubits': each qubit is
manipulated by one player during the `tactics' phase, and then they are independently measured.  Noting
that we can generate any pure initial `strategy' state
$S$ by acting on the zero state
$\ket{00}$ with an appropriate unitary operator ${\hat J}$, we see the similarity to the earlier scheme
shown in Fig 1b. Here we again prepare a 2-qubit system, allow players to make local manipulations of
the qubits, and finally measure the qubits to determine payoffs just as in Fig 1a(ii). The only
fundamental difference is that Eisert {\em et al} employ a second, inverse gate $J^\dagger$. The second
gate ensures that the structure of the classical game remains embedded within the quantum game -- this
is elaborated upon in Ref.\cite{multiPlayer}. However, the terminology used by Eisert {\em et al} is
completely different -- the manipulations applied by the players are referred to as `moves' (rather than
`tactics'), and moreover the term `strategy' refers to a player's overall act of
choosing what move to play. These terms seem to be more consistent with traditional game theory, however
this may be a matter of personal preference -- the important point is that the underlying structures of
the two schemes are very similar. 
\begin{figure}
\centerline{\epsfig{file=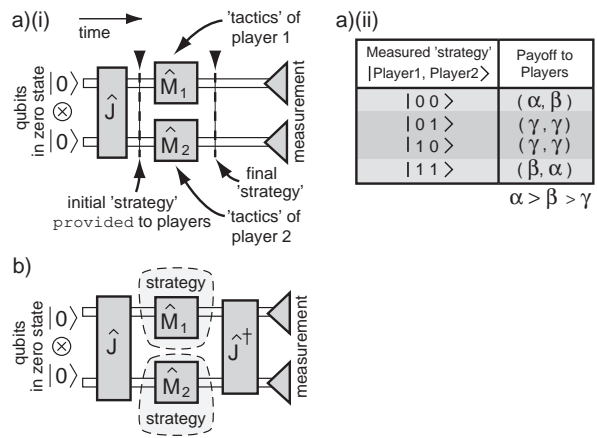,width=7.8cm}}
\vspace{0.2cm}
\caption{a)(i) Our depiction of the scheme proposed by Marinatto and
Weber for quantizing games. The two horizontal lines represent qubits.
a)(ii) the payoff table defining the game, {\em Battle of the Sexes}. In the traditional game, there are
two Nash equilibria, corresponding to both players choosing `0', or both choosing `1'. However without
communication between players they cannot choose which  equilibrium to aim for. This is true even when
$\alpha=\beta$. b) The quantization scheme previously proposed by Eisert {\em et al}.}
\label{figure1}
\end{figure}

We now turn to the question of whether the players of the quantum game can escape from the dilemma
suffered by their traditional counterparts. Marinatto and
Weber claim that this occurs when the initial strategy is set to ${1\over
\sqrt{2}}(\ket{00}+\ket{11})$, and the player's `tactics' are limited to a probabilistic choice between
applying the identity,
${\hat I}$, and applying $\hat{C}=\hat{\sigma}_x$, the Pauli spin-flip operator
(this is a severe restriction on the full range of quantum mechanically possible manipulations). The
symbols
$p^*$ and
$q^*$ are used to represent the probability that ${\hat \sigma_x}$ will be chosen by players 1 and 2,
respectively. The only `tactics' consistent with maximum expected payoff are
$(p^*=0,q^*=0)$, i.e. both players choosing to apply ${\hat I}$ with certainty, or
$(p^*=1,q^*=1)$, i.e. both choosing to apply
${\hat \sigma_x}$. In either case the net effect is that the final `strategy' is unchanged from the
initial form, so that the measurement will yield either $\ket{00}$ or $\ket{11}$ with equal probability.
Hence the expected payoff is
$(\alpha+\beta)/2$ to each player. Now, since both $(p^*=0,q^*=0)$ and $(p^*=1,q^*=1)$ lead to the same
final `strategy', Marinatto and
Weber claim that this is therefore the unique solution in the game. However, it seems to us that a clear
dilemma remains for the players, even given that they have performed the above reasoning -- should they
opt for  ($p^*=0$,$q^*=0$), or for $(p^*=1,q^*=1)$? Without knowing what the other player will
choose, there is no way to make this decision! If the tactics are mismatched, i.e. if either
$(p^*=0,q^*=1)$ or
$(p^*=1,q^*=0)$ are adopted, then the worst-case payoffs occur. This is almost exactly the same
problem faced by the players in the traditional game. However there is an important difference. In the
traditional game, the two solutions have
different payoffs to the players:
$(\alpha,\beta)$ or
$(\beta,\alpha)$. Thus each player prefers a different solution, and even if the players
are allowed to communicate, there is no formal mechanism to decide who should get the high payoff. In
the quantum game, since both solutions payout
$(\alpha+\beta)/2$ to each player, there is no such difficulty, and permitting the players to
communicate immediately allows them to agree on which solution to play for.

It is worth remarking on what would occur if the `tactics' employed by Marinatto and
Weber were less restricted. The following remarks apply when the initial `strategy' supplied to the
players is ${1\over
\sqrt{2}}(\ket{00}+\ket{11})$. If the players were allowed to apply any local unitary operation on their
qubit, i.e. any SU(2) matrix, then the maximum possible payoff to any player would remain
$(\alpha+\beta)/2$. Moreover, for {\em any} given tactic
${\hat A}$ applied by player 1, there is a corresponding tactic for player 2 that will
result no net change to the `strategy', and hence achieve maximum payoff to both players. The enquired
tactic is simply the conjugate of ${\hat A}$, i.e. ${\hat B}={\hat A}^*$. The problem of knowing which
tactic has been adopted by the other player remains, just as discussed above. More interestingly, if the
permitted tactics are extended to include all quantum mechanically possible manipulations, including
measurement, entanglement which ancilla qubits, etc., then the quantum game can become {\em more} like
the classical game! This is because either player can now choose to destroy the entanglement in the
`strategy' state, by simply measuring her qubit, and optionally flipping it. Moreover the players
have a motivation to do this since the maximum possible payoff is then increased from
$(\alpha+\beta)/2$ to
$\alpha$. Therefore, with this most general kind of quantum tactic, even the limited advantage mentioned
at the end of the last paragraph disappears, and the dilemma in the quantum game seems just as difficult
as that in the traditional version.

The author is supported by EPSRC.


\begin{references}
\bibitem{theSame} In fact Marinatto and
Weber specify that the
outcome of the measurement determines a pure classical strategy that the players then proceed to
use in a regular round of the game -- but since this process is deterministic it is equivalent to
rewarding the players immediately.
\end{references}
\end{document}